\documentclass[sn-vancouver,Numbered,pdflatex]{sn-jnl}

\usepackage{graphicx}%
\usepackage{multirow}%
\usepackage{amsmath,amssymb,amsfonts}%
\usepackage{amsthm}%
\usepackage{mathrsfs}%
\usepackage[title]{appendix}%
\usepackage{xcolor}%
\usepackage{textcomp}%
\usepackage{manyfoot}%
\usepackage{booktabs}%
\usepackage{algorithm}%
\usepackage{algorithmicx}%
\usepackage{algpseudocode}%
\usepackage{listings}%

\theoremstyle{remark}
\theoremstyle{definition}

\usepackage{float}
\let\origfigure\figure
\let\endorigfigure\endfigure
\renewenvironment{figure}[1][2] {
    \expandafter\origfigure\expandafter[H]
} {
    \endorigfigure
}
\usepackage{lscape}
\usepackage{pdflscape}
\newcommand{\blandscape}{\begin{landscape}}
\newcommand{\elandscape}{\end{landscape}}
\usepackage[none]{hyphenat}
\usepackage[export]{adjustbox}
\usepackage{multirow}
\usepackage{longtable}
\usepackage{array}
\usepackage{wrapfig}
\usepackage{colortbl}
\usepackage{booktabs} 
\usepackage{colortbl}
\usepackage{pdflscape}
\usepackage{tabu}
\usepackage{threeparttable}
\usepackage[normalem]{ulem}
\usepackage{threeparttablex}
\usepackage{makecell}
\usepackage{array}
\newcolumntype{C}[1]{>{\centering\arraybackslash}p{#1}}
\newcolumntype{M}[1]{>{\centering\arraybackslash}m{#1}}
\newcolumntype{N}{@{}m{0pt}@{}}
\usepackage{caption}
\usepackage{graphicx}
\usepackage{siunitx}
\usepackage{hhline}
\usepackage{calc}
\usepackage{tabularx}

\usepackage{threeparttable}
\usepackage{wrapfig}
\usepackage{adjustbox}
\usepackage{hyperref}

\usepackage{booktabs}
\usepackage{longtable}
\usepackage{array}
\usepackage{multirow}
\usepackage{wrapfig}
\usepackage{float}
\usepackage{colortbl}
\usepackage{pdflscape}
\usepackage{tabu}
\usepackage{threeparttable}
\usepackage{threeparttablex}
\usepackage[normalem]{ulem}
\usepackage{makecell}
\usepackage{xcolor}

\usepackage{longtable,booktabs,array}
\usepackage{calc} 
\usepackage{etoolbox}
\makeatletter
\patchcmd\longtable{\par}{\if@noskipsec\mbox{}\fi\par}{}{}
\makeatother
\IfFileExists{footnotehyper.sty}{\usepackage{footnotehyper}}{\usepackage{footnote}}
\makesavenoteenv{longtable}

\begin{document}

\title[Ethical modelling software]{Ethical considerations when planning, implementing and releasing health economic model software: a new proposal}


\author*[1,2]{\fnm{Matthew} \spfx{P} \sur{Hamilton} }\email{\href{mailto:matthew.hamilton1@monash.edu}{\nolinkurl{matthew.hamilton1@monash.edu}}}

\author[2,4,1]{\fnm{Caroline} \sur{Gao} }

\author[3]{\fnm{Jonathan} \sur{Karnon} }

\author[5]{\fnm{Luis} \sur{Salvador-Carulla} }

\author[2,4]{\fnm{Sue} \spfx{M} \sur{Cotton} }

\author[1]{\fnm{Cathrine} \sur{Mihalopoulos} }

  \affil*[1]{\orgdiv{School of Public Health and Preventive Medicine}, \orgname{Monash University}, \orgaddress{\city{Melbourne}, \country{Australia}}}
  \affil[2]{\orgname{Orygen}, \orgaddress{\city{Parkville}, \country{Australia}}}
  \affil[3]{\orgdiv{Flinders Health and Medical Research Institute}, \orgname{Flinders University}, \orgaddress{\city{Adelaide}, \country{Australia}}}
  \affil[4]{\orgdiv{Centre for Youth Mental Health}, \orgname{The University of Melbourne}, \orgaddress{\city{Parkville}, \country{Australia}}}
  \affil[5]{\orgdiv{Health Research Institute}, \orgname{University of Canberra}, \orgaddress{\city{Canberra}, \country{Australia}}}

\abstract{\textbf{Summary: } Most health economic analyses are undertaken with the aid of computers. However, the research ethics of implementing health economic models as software (or computational health economic models (CHEMs)) are poorly understood. We propose that developers and funders of CHEMs should adhere to research ethics principles and pursue the goals of: (i) socially acceptable user requirements and design specifications; (ii) fit for purpose implementations; and (iii) socially beneficial post-release use. We further propose that a transparent (T), reusable (R) and updatable (U) CHEM is suggestive of a project team that has largely met these goals. We propose six criteria for assessing TRU CHEMs: (T1) software files are publicly available; (T2) developer contributions and judgments on appropriate use are easily identified; (R1) programming practices facilitate independent reuse of model components; (R2) licenses permit reuse and derivative works; (U1) maintenance infrastructure is in place; and (U2) releases are systematically retested and deprecated. Few existing CHEMs would meet all TRU criteria. Addressing these limitations will require the development of new and updated good practice guidelines and investments by governments and other research funders in enabling infrastructure and human capital. \newline \newline}



\maketitle

\hypertarget{introduction}{%
\section{Introduction}\label{introduction}}

Health economics is a discipline concerned with problems that arise due to scarce resources, such as how to value health and healthcare, allocate healthcare budgets and configure health services \citep{wagstaff2012four}. In seeking to solve these problems, health economists typically use models which are simplified and selective representations of systems that are believed to influence human health. A health economic model should be capable of being described using words and figures (a conceptual model), equations (a mathematical model) and software (a computational model). Health economic scientific manuscripts typically involve description of a model using its conceptual and mathematical representations and report results that have been generated by its computational representation (i.e., the format that allows computers to perform calculations). Authors of these manuscripts assume (rarely explicitly) that the conceptual, mathematical and computational representations of a model are internally consistent and externally valid. The feasibility of independently assessing the validity of this assumption depends in part on how a health economic model has been implemented computationally.

The set of software tools and files implementing a health economic model's structure, data and algorithms can be called a computational health economic model (CHEM). CHEMs can be developed using commercial modelling or spreadsheet software (for example, Microsoft Excel \citep{msexcel}) or open-source programming languages such as Python \citep{python2009} or R \citep{RCORE2022}. Training and professional practice guidance for health economists typically provide only superficial coverage of software development topics. However, these capabilities may be important for achieving a number of proposed methodological innovations for advancing the health economics field. Living health technology assessments \citep{thokala2023living}, machine learning powered health economic analyses \citep{padula2022machine}, open source health-economic models \citep{Pouwels2022} and the use of Large Language Models (LLMs) to program health economic models \citep{poirrier2023msr26} all require careful consideration of CHEMs as software development projects.

There are important ethical dimensions specific to software development that the health economics community should engage with more directly. For example, recent developments in LLMs pose novel ethical issues for software developers including: new security risks; more restrictive open-source licensing; environmentally harmful levels of computing power; and concerns about model explainability and unintended consequences \citep{ozkaya2023application}. The ethical dimensions of computational public health modelling remain poorly understood \citep{10.3389/fpubh.2017.00068}. However, documents such as the World Health Organisation's recent guidance on generative artificial intelligence \citep{world2024ethics} can be reference points to help identify ethical topics relevant to health economic model software development.

\begin{table}
\centering
\caption{\label{tab:proscons}Implications of choosing proprietary software or an open-source programming language for the computational implementation of health economic models}
\centering
\begin{tabular}[t]{>{\raggedright\arraybackslash}p{7em}>{\raggedright\arraybackslash}p{20em}>{\raggedright\arraybackslash}p{20em}}
\toprule
Implications & Proprietary modelling or spreadsheet software & Open-source programming language\\
\midrule
 & (+) Will include tools to efficiently accomplish many common modelling tasks. & (-) Initial development is likely to be time and resource intensive.\\

 & (+) Requires limited or no programming skills. & (+) Over medium term, efficiency savings are possible if:\\

 & (-) Requires model developers and users to purchase (in some cases, expensive) software licenses. & (i) artefacts (code and data) from pre-existing models are leveraged; and / or\\

\multirow{-4}{7em}{\raggedright\arraybackslash \textbf{Resources}} &  & (ii)    models are to be maintained and transferred (e.g. to new decision problems / contexts).\\
\cmidrule{1-3}
\addlinespace[0.3em]
\multicolumn{3}{c}{\textbf{Transparency}}\\
\hspace{1em} & (+) Popular tools (e.g. Excel) are readily understood by many. & (+) Easy to publicly share testable source code while selectively restricting access to confidential data.\\

\hspace{1em} & (-) Manual transfer of results to manuscripts through copy and paste or transcription from visual inspection increases risk of (undetected) reporting errors. & (+) Easy to make clear who has developed and tested each part.\\

\hspace{1em} &  & (-) Novel code increases likelihood of (undetected) software errors.\\

\addlinespace[0.3em]
\multicolumn{3}{c}{\textbf{Reusability}}\\
\hspace{1em} & (+) Files will reliably open and execute correctly for years after project is completed. & (-) Potentially fragile – if not maintained / not bundled with all required and correctly versioned dependencies.\\

\hspace{1em} &  & (+) Facilitates modular implementations that make complex model representations more tractable.\\

\hspace{1em} &  & (+) Facilitates customised integration with existing data / decision systems.\\

\addlinespace[0.3em]
\multicolumn{3}{c}{\textbf{Updatability}}\\
\hspace{1em}\multirow{-7}{7em}{\raggedright\arraybackslash \textbf{Ethical}} & (-) Modifications and adaptations, particularly by third parties, may require bespoke permissions and it can be difficult to clearly itemise all changes. & (-) Facilitates collaborative development and maintenance.\\
\bottomrule
\end{tabular}
\end{table}

In this paper, we draw on our experience with developing and using CHEMs and consider literature on modelling and software development practice to propose: (i) an initial set of research ethics goals for CHEM developers and funders; (ii) attributes of CHEMs that suggest achievement of these goals; and (iii) criteria against which these attributes can be assessed. \\

\hypertarget{research-ethics-goals-for-chem-developers-and-funders}{%
\section{Research ethics goals for CHEM developers and funders}\label{research-ethics-goals-for-chem-developers-and-funders}}

Resnik \citep{resnik2020ethics} identified eighteen ethical principles that underpin influential codes of research good practice. We suggest that six of these principles are particularly relevant to CHEM implementation decisions: \emph{accountability} (researchers declaring and taking responsibility for their specific contributions); \emph{confidentiality} (protecting patient records and other confidential information), \emph{intellectual property} (honoring copyrights and acknowledging contributors), \emph{openness} (sharing project data, tools and resources), \emph{transparency} (disclosing the methods and materials needed to evaluate research) and \emph{social responsibility} (promoting social good).

Boden and McKendrick \citep{10.3389/fpubh.2017.00068} propose a framework for ethical public health modelling based on the criteria of \emph{independence} (concerning how modeller subjectivity shapes model design), \emph{beneficence / non-maleficence} (concerning model quality and utility), \emph{transparency} (concerning the need for policymakers to reliably evaluate model strengths and weaknesses) and \emph{justice} (concerning the social obligations of modellers to consider and communicate ethical issues about model use). This framework is not specific to a model's implementation as software, but applies to a modelling project as a whole, including a model's conceptual and mathematical representations. Boden and McKendrick proposed 13 questions to evaluate ethical risk across the four criteria, but of these, only two pertain to computational models: ``is the model code open-source or available on request'' (for transparency) and ``has the model been verified, i.e., does it do what the modeller wants it to do?'' (for beneficence / non-maleficence).

To identify ethical issues that are specific to CHEMs, we considered how Resnick's 18 principles and Boden and McKendrick's four criteria might apply to the distinct phases in a software development project. Software development lifecycle (SDLC) models take different approaches to the inclusion, naming, definition, sequencing and iteration of project phases \citep{ruparelia2010software}. However, SDLC models typically have components that can principally map to the concepts of \emph{planning} (e.g., identification of system / user requirements and design specifications), \emph{implementation} (e.g., development and testing) and \emph{release} (e.g., deployment, system integration, maintenance and support). For simplicity, we propose one overarching ethical goal for each phase. These goals are:
(i) \emph{socially acceptable user requirements and design specifications} (during CHEM planning);
(ii) \emph{fitness for purpose} (of CHEM implementations); and
(iii) \emph{socially beneficial use} (post CHEM release).

Transparency (as a Resnik principle and Boden and McKendrick criterion) and Resnik's social responsibility principle underpin all three of the ethical goals we propose (Table \ref{tab:timelygls}). Additionally: (i) the social acceptability goal incorporates Resnik's accountability, confidentiality and intellectual property principles and Boden and McKendrick's justice and independence criteria;
(ii) the fitness for purpose goal relates to Resnik's accountability principle and Boden and McKendrick's independence and beneficence / non-maleficence criteria; and
(iii) the socially beneficial use goal reflects Resnik's openness principle and Boden and McKendrick's beneficence / non-maleficence and justice criteria.

Misalignment between the values of computational model developers and those of the population groups affected by decisions based on their models presents significant ethical risks \citep{thompson2022escape, duckett2022journey}. The value judgments of health economic modellers are rarely adequately specified \citep{duckett2022journey}. These value judgments influence the assumptions, selection of model features and standards for evidence that shape health economic model projects \citep{HARVARD2020112975}. For example, the process of determining CHEM \emph{user requirements and design specifications} will involve making judgments about the competing claims of proprietary and open-source development platforms (Table \ref{tab:proscons}). An advantage of commercial software is that users often require limited or no programming skills to develop and apply CHEMs. Open-source software programming languages, on the other hand, may facilitate the development of CHEMs that are more transparent, reusable and updatable \citep{incerti2019r, Pouwels2022}. The choice of software development platform can influence how easily a CHEM can be amended by others to reflect alternative value judgments relating to model conceptualisation and mathematical formalisation (Table \ref{tab:timelygls}).

Modellers need to do more than self-certify that their computational model is \emph{fit for purpose} - they also need to provide potential third-party users with the means of assessing its adequacy, appropriateness and usefulness \citep{Erdemir2020, Feenstra2022, thompson2019escape, 10.3389/fpubh.2017.00068}. However, it is common for health economic models to have serious methodological flaws \citep{carletto_zanuzzi_sammarco_russo_2020, WONDER2015467}; insufficient validation \citep{Ghabri2019, kolovos2017model, haji2013model}, poor reproducibility \citep{Jalali2021, McManus2019, Bermejo2017}; and undeclared errors \citep{Radeva2020}. Appropriate computational implementation choices can help address many of these shortcomings, for example by automating and publicly documenting quality assurance checks and providing executable code for review and testing by third parties.

The \emph{social benefit} of developing a CHEM may be limited if it is rarely used, if mis-used or when its acceptability and adequacy rapidly decay. Reuse of CHEMs as components of other models can potentially make model development more efficient \citep{Arnold2010, garcia2021cost}. However, health economic models face challenges related to transferability across jurisdictions \citep{garcia2021cost} that create barriers to reuse. Without ongoing maintenance, a CHEM risks becoming less reliable with time \citep{garcia2021cost} and is at risk of being deployed for purposes for which it is poorly suited \citep{calder2018computational}. Currently, health economic models are rarely implemented computationally in a manner that facilitates routine updates \citep{Sampson2017}, thus limiting the temporal window within which a CHEM can be validly applied.

Our proposed ethical goals are most likely to be met when there is both good professional practice by CHEM developers and adequate resourcing by project funders. The choices CHEM developers make often need to balance ethical considerations with project resource constraints (Table \ref{tab:proscons}). Modelling projects implemented by diligent and conscientious CHEM developers may still be exposed to ethical risks if there are inadequate resources to meaningfully engage important stakeholder groups, prepare code and data artefacts for independent assessment and provide ongoing model maintenance and user support. Currently, it can take ``an extraordinary amount of idealism'' to commit to authoring and maintaining research software \citep{anzt2020environment}. Existing incentive structures for research health economists generally do not promote facilitating peers to reuse their work and funding is rarely adequate to routinely update a model. Economists may therefore make pragmatic judgments that proprietary licensing approaches provide the most feasible pathway to mobilising the resources to support model maintenance. For these reasons, we believe responsibility for achieving our proposed ethical goals is shared between CHEM developers and funders. Reflecting the shared nature of this responsibility, we suggest that in most cases where a modelling project fails to meet our ethical goals, it will be more appropriate to describe these shortcomings as ``ethical risks'' (the language used by Boden and McKendrick) than to use accusatory terms like ``unethical practice.''

\hypertarget{chem-attributes-that-can-reduce-ethical-risk}{%
\section{CHEM attributes that can reduce ethical risk}\label{chem-attributes-that-can-reduce-ethical-risk}}

Our proposed ethical goals for each project phase are easier to state than to measure. However, achievement of these goals may be inferred from measurable attributes of CHEMs. As described in Table \ref{tab:timelygls}, we believe that the developers and funders of CHEMs that are transparent, reusable and updatable (TRU) are more likely to have achieved the ethical goals we propose.

\begin{table}
\centering
\caption{\label{tab:timelygls}Transparent, reusable and updatable (TRU) computational health economic models (CHEMs) are suggestive of ethical modelling practice}
\centering
\begin{tabularx}{14cm}{|l|X|X|X|}
\toprule
  & \textbf{Social acceptability} & \textbf{Fitness for purpose} & \textbf{Beneficial use}\\
\midrule
\textbf{Transparent} & \multicolumn{2}{>{\hsize=\dimexpr 2\hsize+2\tabcolsep+\arrayrulewidth}X|}{\centering Modeller judgments, model features and verification checks can be reviewed by third parties.}  & Decision-makers can understand the strengths and weakness of models before applying them.\\[20pt] \cmidrule(l{7mm}r{7mm}){2-4}

\textbf{Reusable} & \multirow{2}{3cm}{\centering Models can be modified to reflect alternative value judgments.} & Third party use increases likelihood of errors being identified. & Can inform more decision problems, in more contexts with less duplicative effort.\\[20pt] \cmidrule(l{7mm}r{7mm}){3-4}

\textbf{Updatable} & & \multicolumn{2}{>{\hsize=\dimexpr 2\hsize+2\tabcolsep+\arrayrulewidth}X|}{\centering Models can be validly used for longer.}  \\[20pt]
\bottomrule
\end{tabularx}
\end{table}

Taking steps to make computational models \emph{transparent} is an important means of reducing ethical risk \citep{10.3389/fpubh.2017.00068}. Guidance on transparency in health economic modelling recommend that model code and data should be clearly documented \citep{Eddy2012}. These transparency guidelines, published twelve years ago, did not recommend sharing model code and data. However, more recent healthcare modelling guidance does recommend public dissemination of such artefacts \citep{Erdemir2020}. Online repository services such as Zenodo \citep{Zenodo2013} and Dataverse \citep{Dataverse2007} provide persistent storage solutions that generate a Digital Object Identifier (DOI) for each code and data collection. Ensuring that calculations are correct and consistent with model specifications is an essential part of CHEM quality assurance \citep{techver2019}. The extensiveness of such verification checks can be reported using the concept of code coverage \citep{ERICWONG2010188} - the proportion of model code that has been explicitly tested. Testing procedures and results can be publicly shared through online code development services like GitHub \citep{github2007}. GitHub also provides citation tools and can transparently record all individual code contributions to a modelling project over its lifecycle. These tools may be particularly useful in complex modelling projects implemented over longer time-frames with a large and changing group of collaborators, where the nature and extent of individual model authorship contributions can become unclear \citep{thompson2022escape}.

Although making a CHEM's code, data and documentation publicly available is increasingly considered good practice, it is insufficient to ensure that a model is \emph{reusable}. Writing model algorithms as collections of functions (short, self-contained and reusable software routines that each perform a discrete task) has been recommended as good scientific computing practice \citep{Wilson_2017} and promotes reuse of model sub-components. Computational implementations that store model code and data in distinct files and locations (as opposed to embedding data such as parameter values into source code) are also easier to selectively modify. Modular implementations enable models to be constructed from replaceable sub-models (modules) \citep{pan2021modular} that can share inputs and outputs with each other or be run as independent models \citep{barros2023empowering}. Modularity can make it easier to selectively replace or update specific parts of a model and to scale up or down the level of granularity \citep{pan2021modular}. Granting permissions to others to use, test and adapt models and their components, can be facilitated by two broad categories of open-source licenses. Some guidance strongly recommends the use of permissive licensing \citep{Wilson_2017} that provides users with great flexibility as to the purposes (including commercial) for which content can be re-used. Alternatively, copyleft licenses \citep{copyleft2022} can be used to require that derivative works created by content users remain open-source.

Models should be \emph{updatable} so that they remain valid for longer, evolving as new evidence emerges and the systems being modelled change \citep{Jenkins2021, garcia2021cost}. Ensuring that a model is regularly reviewed to identify and implement required improvements is a recommended defence against model validity decay \citep{calder2018computational}. Sustainable maintenance of open-source research software requires both a core development team and an active user community \citep{info:doi/10.2196/20028}. Online communities can be an efficient means of engaging model users in testing each version of a model, identifying issues and suggesting improvements. Services such as GitHub \citep{github2007} provide collaborative code development tools \citep{MERGEL2015464} that help elicit, integrate and reconcile contributions from multiple contributors and to ensure each update is uniquely identifiable and retrievable. Verification checks should be rerun with each model update, a task that can be automated using the software development practice of continuous integration \citep{CI2017}. The risk of software revisions having unintended consequences for third party users can be mitigated through the use of deprecation conventions \citep{zhou2016api} that take an informative and staged approach to retiring outdated CHEMs.

\hypertarget{criteria-for-assessing-tru-attributes}{%
\section{Criteria for assessing TRU attributes}\label{criteria-for-assessing-tru-attributes}}

We suggest six criteria (two for each attribute) for assessing the TRU attributes of CHEMs. These criteria are summarised in Table \ref{tab:trucrit}.

\renewcommand{\arraystretch}{2}
\begin{table}
\centering
\caption{\label{tab:trucrit}Criteria for assessing the transparency, reusability and updatability of CHEMs}
\centering
\begin{tabular}[t]{>{\raggedright\arraybackslash}p{8em}>{\raggedright\arraybackslash}p{3em}>{\raggedright\arraybackslash}p{28em}}
\toprule
Attribute & Item & Criterion\\
\midrule
 & T1 & Software files (executable code, testing procedures and outcomes and non-confidential data) are publicly available.\\

\multirow{-2}{8em}{\raggedright\arraybackslash \textbf{Transparent}} & T2 & It is easy to see who funded, developed and tested each part of the CHEM and to identify the modelling team’s assumptions, judgments and theories about CHEM development and use.\\
\cmidrule{1-3}
 & R1 & Programming practices facilitate independent reuse of model components.\\

\multirow{-2}{8em}{\raggedright\arraybackslash \textbf{Reusable}} & R2 & Licenses allow anyone to ethically reuse model code and non-confidential data, in whole or in part, without charge, and for purposes that include the creation of derivative works.\\
\cmidrule{1-3}
 & U1 & Maintenance infrastructure is in place to support version control and collaboration with model users.\\

\multirow{-2}{8em}{\raggedright\arraybackslash \textbf{Updtatable}} & U2 & Each new release of a CHEM is systematically retested, with changes implemented to minimise disruptions for existing model users.\\
\bottomrule
\end{tabular}
\end{table}

\hypertarget{further-issues}{%
\section{Further issues}\label{further-issues}}

Health economic models routinely have limitations in transparency \citep{Jalali2021, McManus2019, Bermejo2017, Ghabri2019}, reusability \citep{Feenstra2022, Emerson2019}, and updatability \citep{Sampson2017, 10.3389/fpubh.2022.899874}. Few current CHEMs would meet all six TRU criteria that we propose, which collectively require that a CHEM is implemented as an open-source project. Despite in principle support from many in the health economics field \citep{Pouwels2022}, open-source CHEMs remain rare \citep{Feenstra2022, Emerson2019}.

Barriers to open-source CHEMs include health economist concerns about intellectual property, confidential data, the risk of model misuse and the resources required for model maintenance \citep{Pouwels2022, Wu2019}. Commercial considerations, for example when health economic models are owned by pharmaceutical companies and consultancies, may also limit the public availability of models and their constituent code and data \citep{Feenstra2022}. Other barriers relate to the software platforms in which CHEMs are authored \citep{Pouwels2022}.

Adherence to good practice guidance is an essential requirement for healthcare modelling \citep{Erdemir2020}, but guidelines for implementing open source CHEMs remain scarce, piecemeal and need improving \citep{Sampson2019}. Training curricula for current and future health economists that develop foundational software development skills have been recommended \citep{kindilien2021role}. Health economists also need to be able to access reliable and up to date legal advice about the implications of alternative approaches to software licensing. The ethical dimensions of open-source software development are broader than the issues we have discussed here and include issues relating to social contracts, contributor autonomy and public goods \citep{grodzinsky2003ethical}.

Achieving the ethical goals we propose will require significant investments by governments and other research funders in enabling infrastructure and human capital. Related investment strategies that have previously been recommended for development of the health economic research field include support for harmonised ethical standards for model development \citep{10.3389/fpubh.2017.00068}, methodological innovation to improve model transferability \citep{craig2018taking}, networks of modellers working on common health conditions \citep{Sampson2019}, and centralized infrastructure such as open-source model repositories \citep{Pouwels2022} and a standard platform for model implementations \citep{Ghabri2019}.

\hypertarget{conclusion}{%
\section{Conclusion}\label{conclusion}}

We propose that CHEM developers and funders adopt the ethical goals of socially acceptable user requirements and design specifications, fit for purpose implementations and socially beneficial post-release use. We suggest that CHEM attributes of transparency, reusability and updatability are indicative of projects that meet these goals and have identified six criteria for assessing these attributes. We hope our proposals contribute to a timely and much needed debate within the health economics community about the ethical dimensions of implementing health economic models as software.

\hypertarget{ethics-approval}{%
\subsection*{Ethics approval}\label{ethics-approval}}
\addcontentsline{toc}{subsection}{Ethics approval}

None.

\hypertarget{funding}{%
\subsection*{Funding}\label{funding}}
\addcontentsline{toc}{subsection}{Funding}

Funded by an Australian Government Research Training Program (RTP) Scholarship to Matthew Hamilton.

\hypertarget{conflict-of-interest}{%
\subsection*{Conflict of Interest}\label{conflict-of-interest}}
\addcontentsline{toc}{subsection}{Conflict of Interest}

None declared.

\newpage
\appendix
\counterwithin{figure}{section}
\counterwithin{table}{section}

\bibliography{Submission.bib}

\end{document}